\documentclass{article} 
\usepackage{lmrl2025_workshop,times}
\usepackage{xcolor}
\usepackage{graphicx}
\usepackage{subcaption}
\usepackage{float}

\usepackage{amsmath,amsfonts,bm}

\def\eqref#1{equation~\ref{#1}}

\def\1{\bm{1}}

\DeclareMathAlphabet{\mathsfit}{\encodingdefault}{\sfdefault}{m}{sl}
\SetMathAlphabet{\mathsfit}{bold}{\encodingdefault}{\sfdefault}{bx}{n}

\usepackage{hyperref}
\usepackage{url}

\title{Identifying Critical Phases for Disease Onset with Sparse Haematological Biomarkers}

\author{
\textbf{Andrea Zerio}\textsuperscript{1}, 
\textbf{Maya Bechler-Speicher}\textsuperscript{2,3}, 
\textbf{Tine Jess}\textsuperscript{1,4} \&
\textbf{Aleksejs Sazonovs}\textsuperscript{1} \\ 
\textsuperscript{1} Center of Excellence for Molecular Prediction of Inflammatory Bowel Disease, PREDICT, \\
Department of Clinical Medicine, Aalborg University \\ 
\textsuperscript{2} Blavatnik School of Computer Science, Tel-Aviv University \\
\textsuperscript{3} Meta \\
\textsuperscript{4} Department of Gastroenterology \& Hepatology, Aalborg University Hospital\\
\texttt{\{anze, jess, alesaz\}@dcm.aau.dk} \quad \texttt{mayab4@mail.tau.ac.il}
}

\lmrlfinalcopy 
\begin{document}

\maketitle

\section{Temporally Sparse Biomarker Data}

Haematological biomarkers from clinical chemistry tests are widely used in medical practice, generating large-scale molecular data that can support health and disease research \citep{uttley2016building, foy2025haematological}. Many of these biomarker values and their dynamics are known to be strong indicators of health-related traits. Emerging evidence indicates that many complex diseases, such as immune-mediated diseases (IMIDs), exhibit pre-diagnostic stages that can be inferred from these biomarkers \citep{vestergaard2023characterizing, deane2010preclinical}. A pre-diagnostic stage refers to a phase where a patient has not yet met the clinical criteria for diagnosis, yet subtle, systemic changes in their biomarker profiles suggest an elevated risk of disease onset. 

Detecting early dysregulation in biomarker patterns is crucial for enabling timely preventative interventions. Unfortunately, routine clinical sampling is guided by medical needs rather than standardized research protocols, introducing confounding noise. As such, similar biomarker values may be observed across multiple conditions, making it difficult to distinguish disease-specific patterns and reducing predictive specificity. More importantly, it results in irregular sampling intervals, introducing sparsity into the data's temporal dimension. All this makes it difficult to apply standard time-series models, which often rely on interpolation or imputation to fill in missing data \citep{herbers2021deal, tanvir2023incomplete}. Such preprocessing can obscure true biological signals, distorting learning patterns, reducing predictive accuracy, and compromising interpretability.

\section{Detecting Disease Onset from Sparse Biomarkers with Interpretable Graph Learning}

Our ultimate goal is to identify biomarker dysregulation periods predictive of disease onset. To address the sampling challenges, we model biomarker trajectories as time-weighted directed graphs, preserving temporal structure without imputation or zero-inflation. In this framework, detecting dysregulation reduces to identifying the key nodes that correspond to critical periods. 

For each individual, we define a graph composed of multiple longitudinal trajectories, one for each of the biomarkers measured throughout their history. These trajectories are represented as directed line graphs \( G_k = (V, E) \), where \( V = \{v_1, v_2, \dots, v_T\} \) is the set of nodes, with each node \( v_t \) representing a sampling event at time \( t \), and where \( E = \{(v_1, v_2), (v_2, v_3), \dots, (v_{T-1}, v_T)\} \) is the set of edges. 
To encode the temporal structure of the data, each edge \( (v_t, v_{t+1}) \in E \) is assigned a weight \( w_t = \rho(\Delta_t) \), computed as a function of the time interval \( \Delta_t = t_{t+1} - t_t \) between consecutive sampling events, where \( \rho \) is a weighting function that maps the time interval \( \Delta_t \) to a scalar.

Since our problem formulation aims to detect important nodes, we leverage and extend the recently proposed interpretable GNAN \citep{bechler2024intelligible}. Unlike black-box deep learning models, GNANs provide intrinsically interpretable predictions by constraining the use of feature cross-products and graph topology, resulting in a transparent architecture which provides node and feature importance metrics. This allows us to trace which biomarkers and time points contribute most to a classification decision, shifting the focus to explaining when and how disease-related changes emerge. We extend the original GNAN formulation to generate node representations as shown in Figure \ref{eq:1}:
\begin{equation} \label{eq:1}
[\mathbf{h}_i]_k = \sum_{j \in V} \rho\left(\frac{1}{0.1 + \Delta t_{ji}}\right) f_k(\mathbf{x}_j^{(k)}),
\end{equation}
where $[\mathbf{h}_i]_k$ is the \textit{k}-th entry of the representation of node $i$, denoted as $\mathbf{h}_i$.
This formulation simplifies the distance function of the original GNAN by leveraging symmetries to make the computation of node distances efficient. Additionally, we incorporate a one-dimensional representation of the biomarker’s one-hot encoding to capture categorical information alongside continuous trajectories. For the full formulation see Appendix \ref{FGNAN_Form}.

We demonstrate our approach using binary classification of the pre-diagnostic trajectories of patients and age-matched controls. We select a set of 2,500 Crohn’s Disease (CD) patients and 2,500 controls, sampling the trajectories of three routine clinical biomarkers: haemoglobin, albumin, and C-reactive protein. These biomarkers were chosen for their widespread use in routine clinical testing and their statistical association with the pre-CD disease state \citep{vestergaard2023characterizing}.

\begin{figure}
\vspace{-5pt}
    \centering
    \begin{subfigure}{0.42\textwidth}
        \centering
        \includegraphics[width=\textwidth]{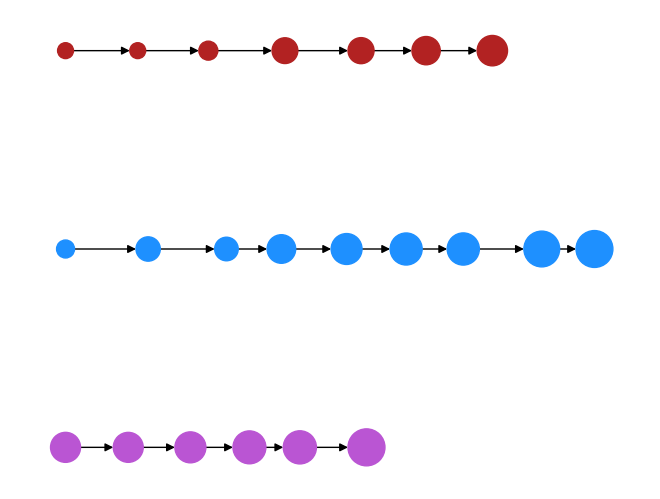}
    \end{subfigure}
    \hfill
    \begin{subfigure}{0.43\textwidth}
        \centering
        \includegraphics[width=\textwidth]{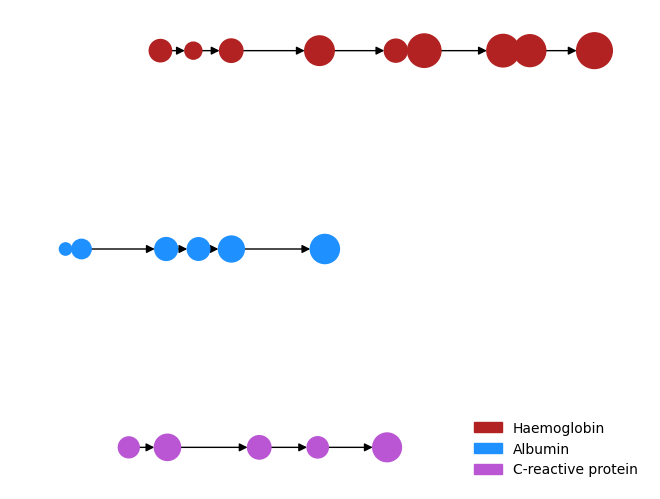}
    \end{subfigure}
    \caption{Diagram illustrating node-level interpretability of synthetically generated data. The size of each biomarker node corresponds to the importance that the model assigned to it. Individual-level trajectories cannot be provided due to Danish data protection laws.}
    \label{fig:node-level-plots}
\end{figure}

As a work-in-progress, the model’s current performance is not yet sufficient for practical application, indicating the need for further development and optimization. However, early results suggest that the model is learning to capture at least some meaningful signals, as shown in Appendix \ref{train_loss}. In order to understand whether the representations of the graphs that the model learns are biologically coherent, we freeze the model and compute the importance of each node in patients' graphs after a single forward pass. To comply with privacy and data-sharing regulations we provide diagrams from synthetic data of node-level interpretability plots in Figure \ref{fig:node-level-plots}.

Our initial results suggest that the model assigns varying importance to nodes, potentially distinguishing between biomarkers and across time. Notably, there is anecdotal evidence that it prioritizes clusters near the trajectory’s end, implying that recent biomarker dynamics could matter more than distant history. This aligns with previous findings that the association strength of the pre-diagnostic signal peaks in the 1–2 years preceding diagnosis \citep{vestergaard2023characterizing}.

\section{Conclusion}

We introduce a novel GNAN-based representation for sparse, temporal biomarker trajectories. Our approach, still in early development, shows promise in learning sparse representations of haematological biomarkers and providing insights into node-level feature importance. This is crucial for clinical and biological applications, where interpretability and efficient representation are key for decision-making and detecting early disease signs. Future work will explore extending the GNAN model further by using recurrent architectures to model temporal nodes, expand support for a broader set of biomarkers, integrate non-biomarker features (e.g., comorbidities), and model interactions across multiple omics layers.

\newpage
\subsubsection*{Meaningfulness Statement}

Understanding health and disease requires models that extract biologically relevant patterns from complex, sparse, and available data. Patient medical history contains an incomplete yet invaluable imprint of one of the most important aspects of life: our health. In this work, we propose using data from routine blood tests, something that each of us has experienced in our lifetime, to learn biomarker trajectories representative of our health. While we demonstrate our approach on CD, its potential extends far beyond, offering a versatile framework for uncovering critical health patterns in any longitudinal setting.

\bibliography{iclr2025_conference}
\bibliographystyle{iclr2025_conference}
\newpage
\appendix
\section{Appendix}
\subsection{Time-delta GNAN Formulation}
\label{FGNAN_Form}
Given a node feature vector consisting of \( d \) individual features, each feature is treated as a separate univariate series and processed by its corresponding function \( f_k \). For node \( i \), the feature representation is:
\begin{equation}
    [\mathbf{h}_i]_k = \sum_{j \in V} \rho\left(\frac{1}{0.1 + \Delta t_{ji}}\right) f_k(\mathbf{x}_j^{(k)}),
\end{equation}

where \( \rho \) is a distance weighting function and \( f_k \) processes the corresponding feature set \( S_k \). 

This simplifies the original GNAN formulation in two ways:
\begin{enumerate}
    \item The original $\#\text{dist}_i(j)$ function is replaced with $\Delta t_{ji}$, which is computationally more efficient to calculate.
    \item Since $\Delta t_{ji}$ represents the difference in time between two dates in a directed graph, every valid path between two nodes is equal in length. As such, we can remove the normalisation term $\frac{1}{\#\text{dist}_i(j)}$.
\end{enumerate}

In addition to processing each biomarker feature independently, we introduce an additional one-dimensional, node-level feature derived from the one-hot encoding of the biomarker. This feature is computed using a multivariate function \( F_{\text{oh}} \), which takes as input the one-hot encoded representation of the biomarker across nodes and aggregates it accordingly. Specifically, for node \( i \), this additional feature is given by:
\begin{equation}
[\mathbf{h}_i]_{\text{oh}} = \sum_{j \in V} \rho\left(\frac{1}{0.1 + \Delta t_{ji}}\right) F_{\text{oh}}(\mathbf{x}_j^{\text{(oh)}}),
\end{equation}

where \( \mathbf{x}_j^{\text{(oh)}} \) represents the one-hot encoding of the biomarker at node \( j \). This ensures that GNAN captures categorical biomarker information alongside continuous biomarker trajectories.

The final node representation is:
\begin{equation}
\mathbf{h}_i = \big([\mathbf{h}_i]_1, [\mathbf{h}_i]_2, \dots, [\mathbf{h}_i]_d, [\mathbf{h}_i]_{\text{oh}}\big).
\end{equation}

For graph-level tasks, we apply sum pooling:
\begin{equation}
\mathbf{h}_G = \sum_{i \in V} \mathbf{h}_i,
\end{equation}
followed by a readout function for prediction:
\begin{equation}
\sigma\left(\sum_{k=1}^d [\mathbf{h}_G]_k\right),
\end{equation}

where \( \sigma \) is an activation function (e.g., sigmoid for classification). Importantly, GNAN enables interpretability by quantifying the influence of node \( j \) on feature set \( S_k \):
\begin{equation}
\text{Influence}(j, S_k, G) = f_k(\mathbf{x}_j^{(k)}) \sum_{i \in V} \rho\left(\frac{1}{0.1 + \Delta t_{ji}}\right),
\end{equation}

allowing us to identify critical time points that contribute most to the prediction. The total contribution of node \( i \) to the graph-level decision is:
\begin{equation}
\text{TotalContribution}(i) = \sum_{j \in V} \rho\left(\frac{1}{0.1 + \Delta t_{ji}}\right) \sum_{k=1}^d f_k(\mathbf{x}_j^{(k)}).
\end{equation}

This formulation ensures that GNAN not only models sparse biomarker trajectories effectively, but also provides an interpretable framework for identifying critical phases in disease progression.

\subsection{Experimental Details}
\label{exp_details}
\subsubsection{Model Training}
We trained a series of FeatureGroupGNAN models, all with ReLU activations, layers in the \{3,\nobreak5\} and hidden channels in the \{100, 64\} range. We used an Adam optimizer over 10 epochs with a custom CosineAnnealingWarmRestartsDecay scheduler, with T0 in the \{10, 50\} range, decay factor in the \{0.3, 0.8\} range, learning rate in the \{1e-5, 5e-5\} range and minimum learning rate in the \{5e-8, 5e-9\} range.

All models were trained on a single NVIDIA Tesla V100-PCIE-16GB GPU.

\subsubsection{Dataset}
\begin{figure} [H]
    \centering
    \begin{subfigure}{\textwidth}
        \centering
        \includegraphics[width=\textwidth]{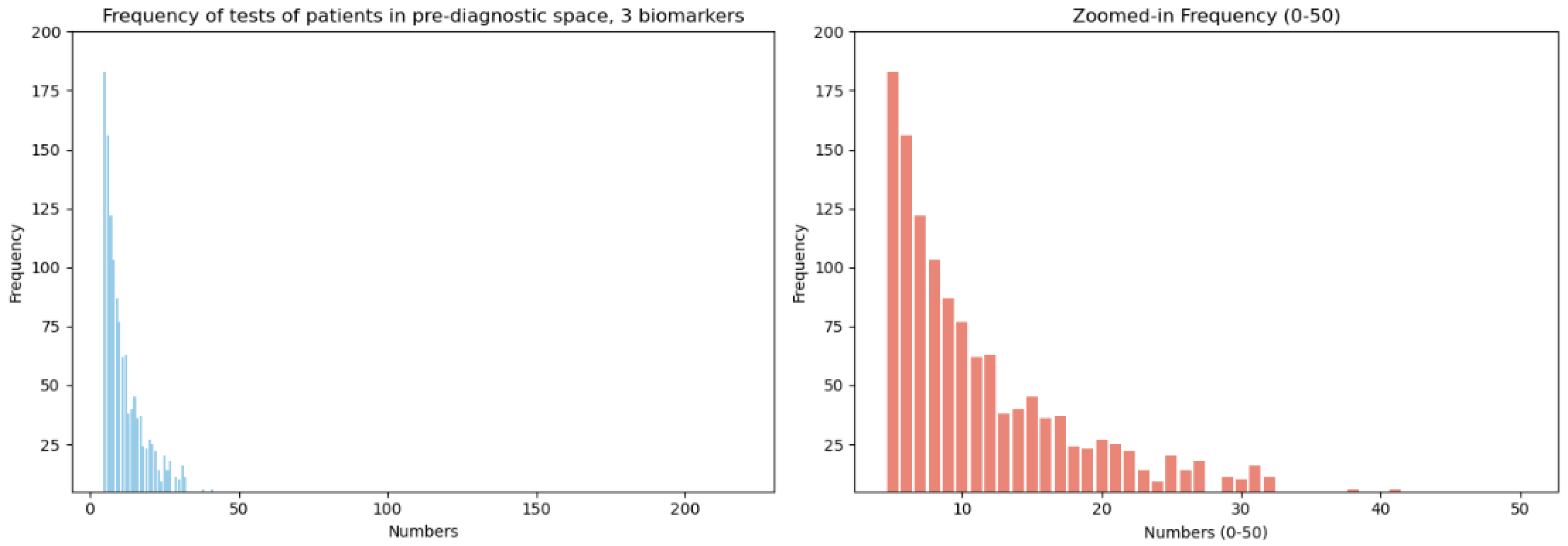}
        \caption{Patient Data Distribution}
    \end{subfigure}
    
    \vspace{0.2cm} 
    
    \begin{subfigure}{\textwidth}
        \centering
        \includegraphics[width=\textwidth]{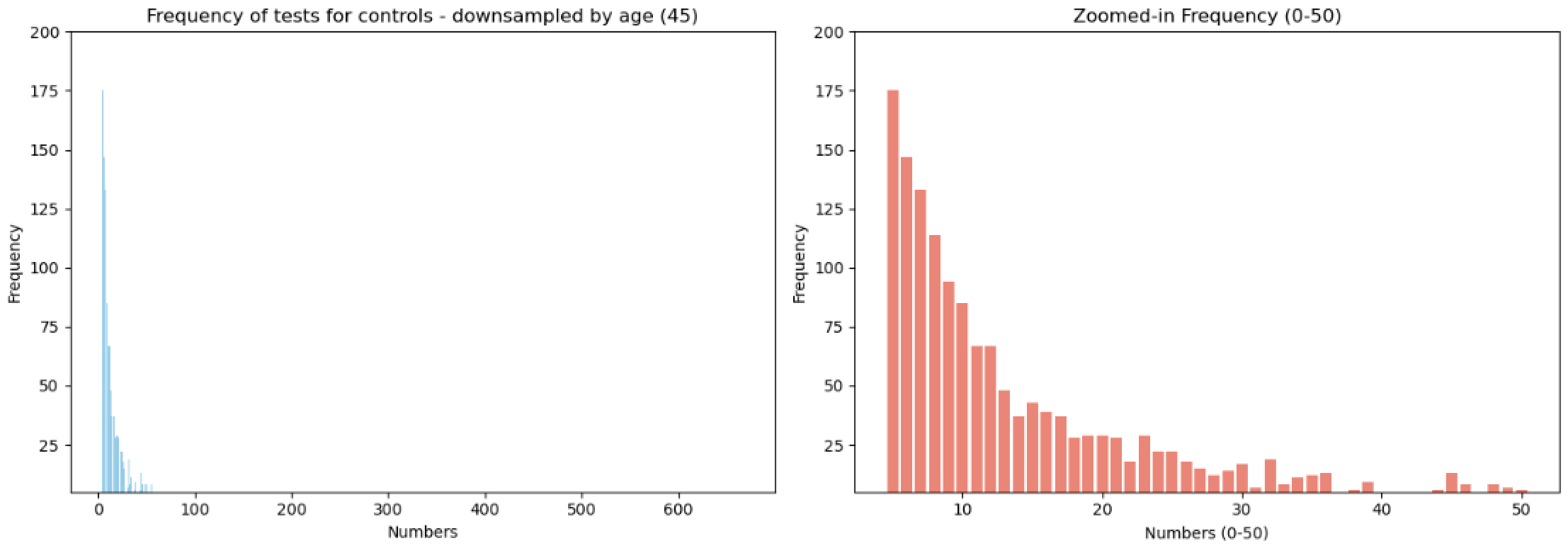}
        \caption{Control Data Distribution}
    \end{subfigure}
    
    \caption{Comparison of patient and control data distributions after downsampling controls by age. Values below n=5 were excluded due to Danish data protection rules.}
    \label{fig:distributional_matching}
\end{figure}

We train and experiment on a dataset of 2,500 individuals who eventually develop CD and 2,500 controls. The patients are sampled at random from a larger nation-wide cohort of CD patients. Similarly the controls are sampled at random from a cohort of around 9 million individuals who were never diagnosed with CD. All data referenced has been obtained from the Danish healthcare registries. We select only the pre-diagnostic trajectories of patients, meaning the biomarkers that were sampled before a formal diagnosis. We then downsampled blood tests from controls to align with the typical age of onset in CD, ensuring a comparable testing frequency between controls and patients. The resulting distributions over the number of tests per-person is displayed in Figure \ref{fig:distributional_matching}. To comply with privacy and data-sharing regulation we exclude all samples with less than 5 measurements before computing the adjusted distributions and producing the plots.

\subsection{Initial Performance Details}
\label{train_loss}
Although still in early development, our model achieves performance comparable to baseline methods reported in UK Biobank (UKBB) studies for similar biomarkers \citep{sazonovs2025op19}. However, a key distinction is that these baseline models were evaluated on data with homogeneous time-point sampling, whereas our model was developed and tested on temporally sparse trajectories. Furthermore, UKBB data is subject to higher informational bias, as control participants were generally healthier due to the requirement to attend one of the recruitment facilities. In contrast, control data in our Danish cohort was hospital-sampled, likely representing a population with a worse overall health profile.

\begin{figure}[H] 
    \centering
    \begin{subfigure}{\textwidth}
        \centering
        \includegraphics[width=0.7\textwidth]{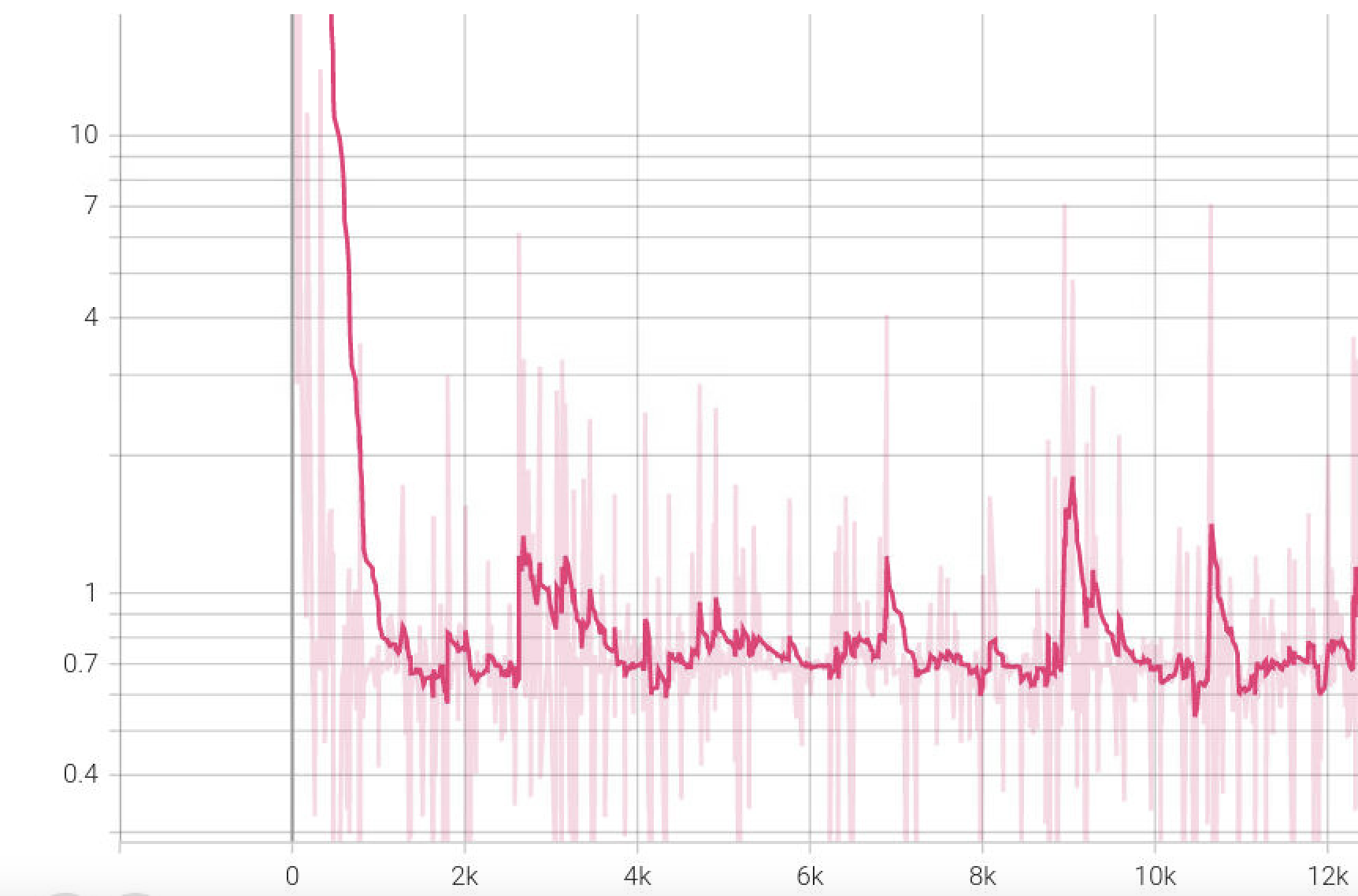}
        \caption{Higher learning rate decay and faster restart}
        \label{fig:loss1}
    \end{subfigure}
    
    \vspace{0.3cm} 
    
    \begin{subfigure}{\textwidth}
        \centering
        \includegraphics[width=0.7\textwidth]{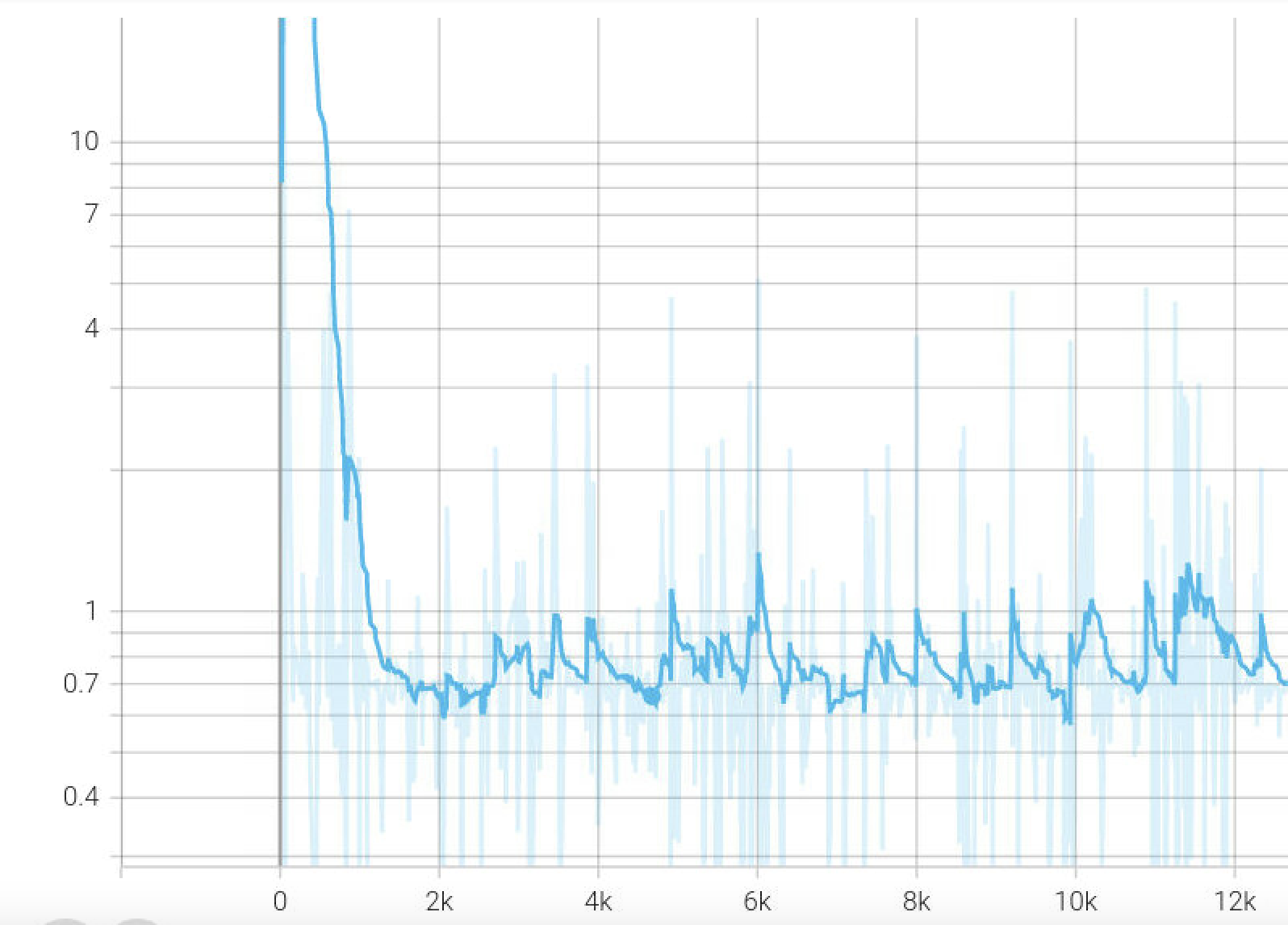}
        \caption{Lower learning rate decay and slower restart}
        \label{fig:loss2}
    \end{subfigure}
    
    \caption{Logspace of batch training loss (BCEWithLogitsLoss). The plots demonstrate the model is capable of learning some initial signal, before finding a local minimum early on and stabilising.}
    \label{fig:loss_plots}
\end{figure}

\begin{figure}
    \centering
    \begin{subfigure}{\textwidth}
        \centering
        \includegraphics[width=1.0\textwidth]{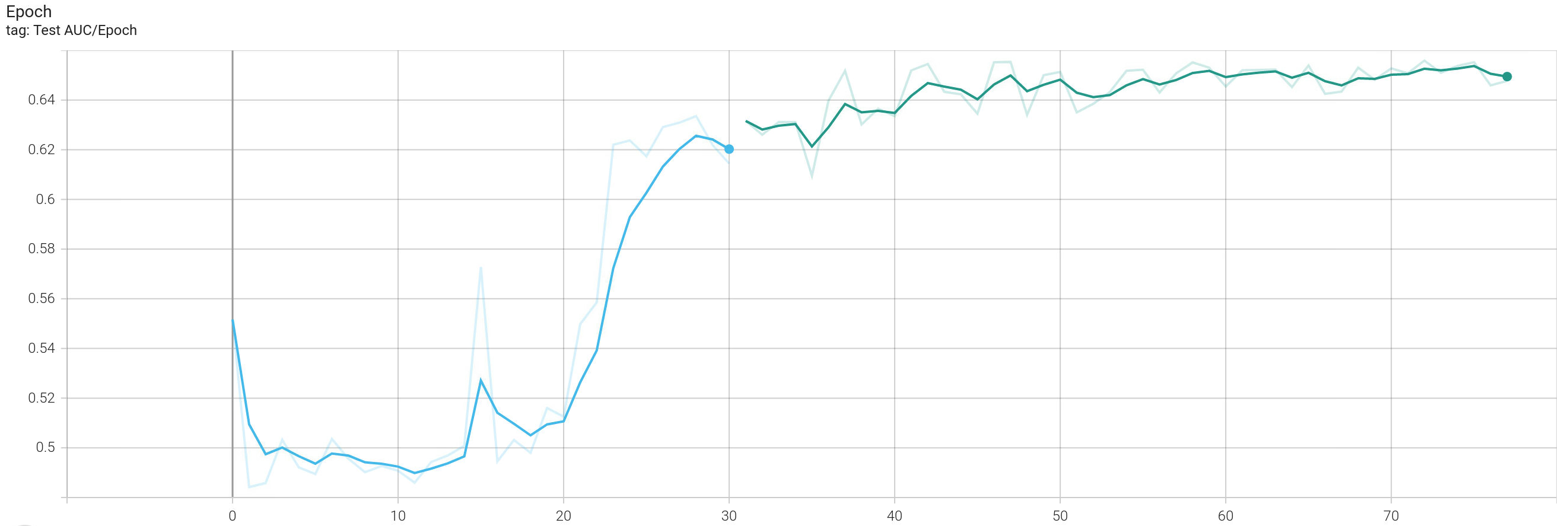}
        \caption{Test Area under the curve (AUC)}
        \label{fig:auc}
    \end{subfigure}
    
    \vspace{0.3cm} 
    
    \begin{subfigure}{\textwidth}
        \centering
        \includegraphics[width=1.0\textwidth]{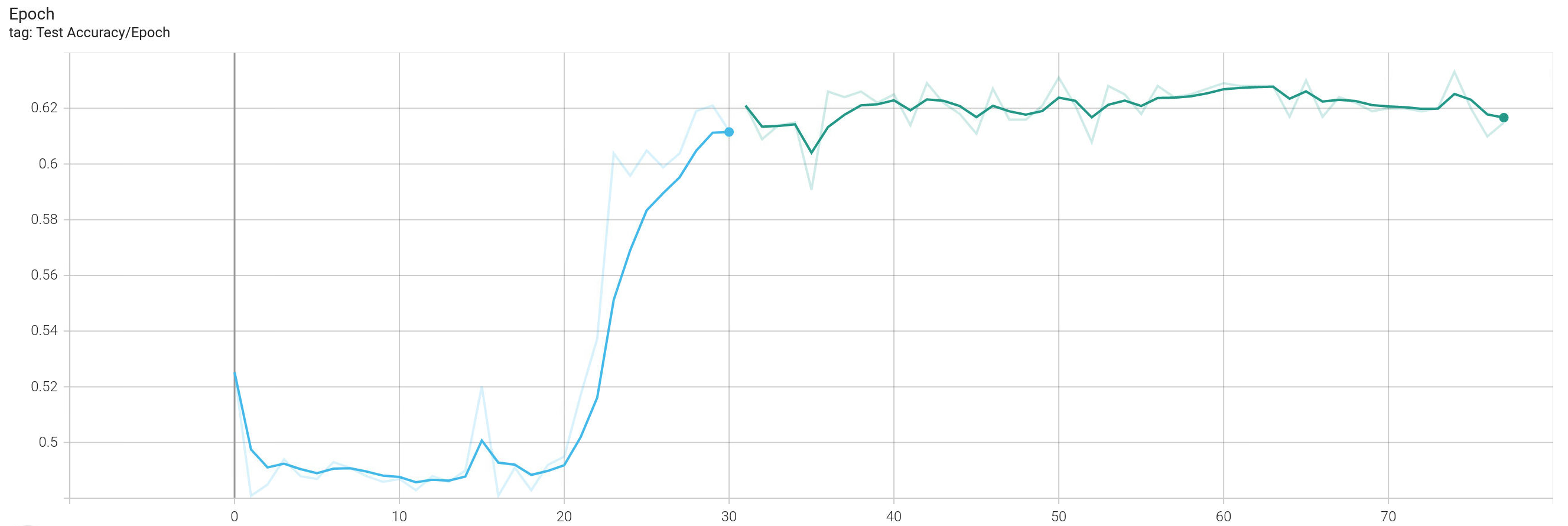}
        \caption{Test Accuracy}
        \label{fig:acc}
    \end{subfigure}
    
    \caption{Test AUC and Accuracy across training. The two lines in each plot correspond to different segments of the same training run, where training was resumed from a checkpoint after reaching the initial stopping point}
    \label{fig:auc_acc}
\end{figure}

\begin{figure}
    \centering
    \begin{subfigure}{0.48\textwidth}
        \centering
        \includegraphics[width=\textwidth]{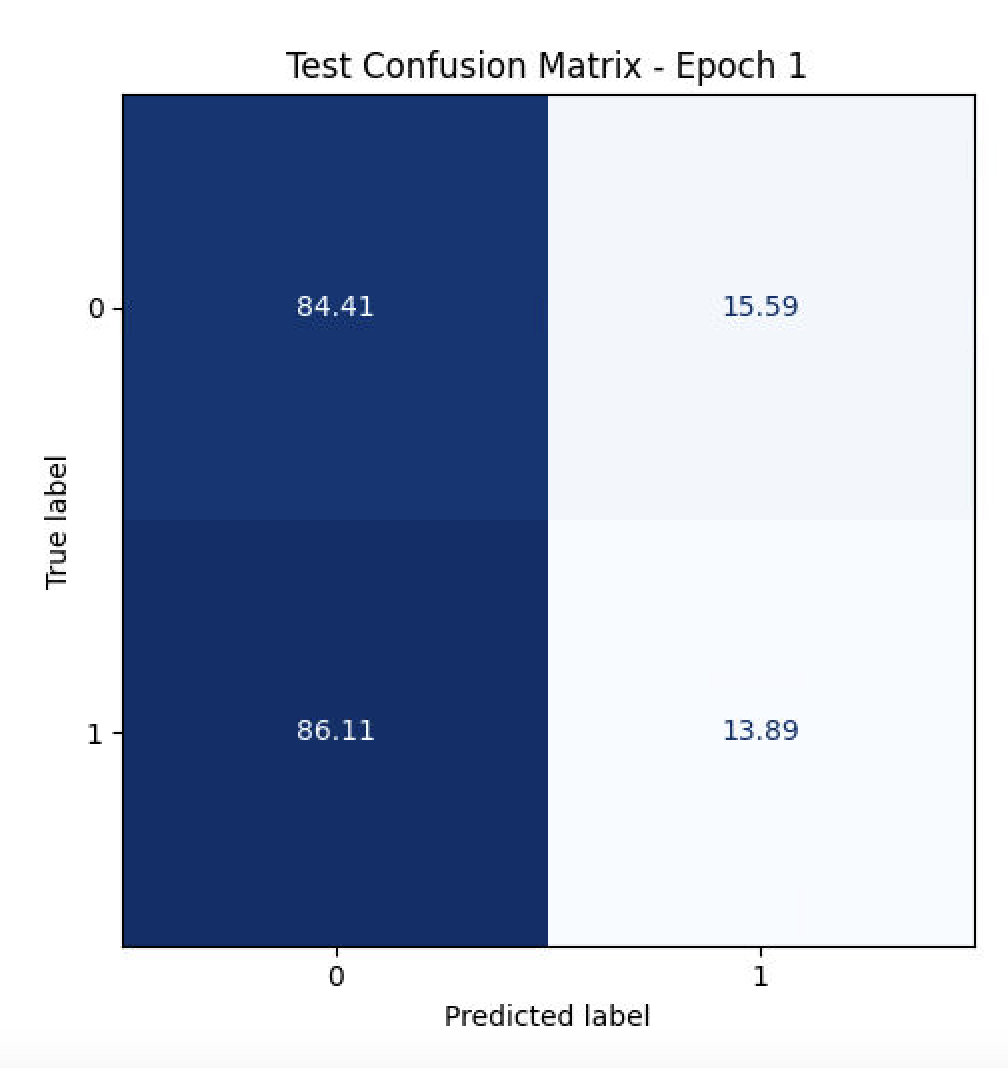}
        \caption{Test Confusion Matrix after 1 epoch}
    \end{subfigure}
    \hfill
    \begin{subfigure}{0.48\textwidth}
        \centering
        \includegraphics[width=\textwidth]{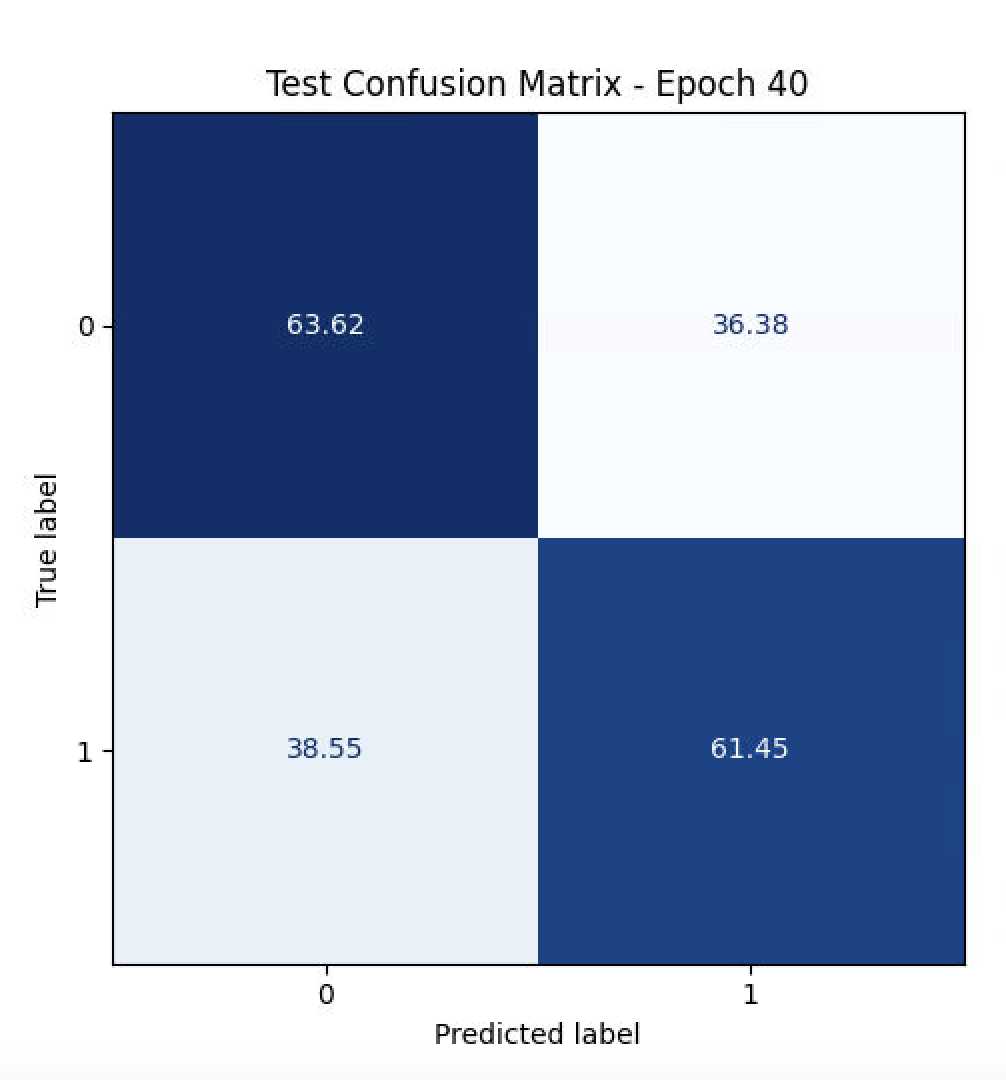}
        \caption{Test Confusion Matrix after 40 epochs}
    \end{subfigure}
    \caption{Comparison of test set confusion matrices. Despite not yet being performant enough to be clinically relevant, the model does seem to learn some initial signal that is discriminative of CD patients and controls.}
    \label{fig:conf-matrix}
\end{figure}
\newpage
\subsection{Funding and Acknowledgement}
The work was supported by grant DNRF148 from the Danish National Research Foundation Center of Excellence and grant NNF23OC0087616 from the Novo Nordisk Foundation. We would like to thank Marie Vibeke Vestergaard for her input and advice with regards to this manuscript.
\end{document}